# Epitaxial Graphene Electronic Structure And Transport

Walt A. de Heer[1], Claire Berger[1,2], Xiaosong Wu[1], Mike Sprinkle[1], Yike Hu[1], Ming Ruan[1], Joseph A. Stroscio[3], Phillip N. First[1], Robert Haddon[4], Benjamin Piot[5], Clément Faugeras[5], Marek Potemski[5], Jeong-Sun Moon[6]

[1] School of Physics, Georgia Institute of Technology, Atlanta, GA-30332, USA
[2] Institut Néel - CNRS, Grenoble, 38042 Cedex 9, France
[3] Center for Nanoscale Science and Technology, NIST, Gaithersburg MD, 20899
[4] Center for Nanoscale Science and Engineering, Departments of Chemistry and Chemical & Environmental Engineering, University of California, Riverside, California 92521
[5] LNCMI -CNRS, Grenoble, 38042 Cedex 9, France
[6] HRL Laboratories LLC, Malibu, CA 90265, USA
email: walt.deheer@physics.gateh.edu

**Abstract:** Since its inception in 2001, the science and technology of epitaxial graphene on hexagonal silicon carbide has matured into a major international effort and is poised to become the first carbon electronics platform. A historical perspective is presented and the unique electronic properties of single and multilayered epitaxial graphenes on electronics grade silicon carbide are reviewed. Early results on transport and the field effect in Si-face grown graphene monolayers provided proof-of-principle demonstrations. Besides monolayer epitaxial graphene, attention is given to C-face grown multilayer graphene, which consists of electronically decoupled graphene sheets. Production, structure, and electronic structure are reviewed. The electronic properties, interrogated using a wide variety of surface, electrical and optical probes, are discussed. An overview is given of recent developments of several device prototypes including resistance standards based on epitaxial graphene quantum Hall devices and new ultrahigh frequency analog epitaxial graphene amplifiers.

## 1. Introduction

Epitaxial graphene (*1*) is rapidly becoming the strongest candidate for post-CMOS electronics and the first commercial devices are actually on the horizon. It is interesting to note that carbon-based electronics has long been recognized as a viable replacement for silicon (*2*). Molecular electronics was the first alternative to be considered, but it never developed beyond rudimentary prototypes(*3*). One persistent problem was in contacting and interconnecting molecules. The unacceptably large contact resistances between the molecule and the metallic contacts remains an unsolved problem. Carbon nanotube electronics provided an important step forward in carbon-based electronics(*4*). Carbon nanotubes are basically large organic molecules that reveal quantum confinement effects and ballistic conduction (*5*), which are both essential features for an electronics platform that can succeed silicon. In that sense nanotube-based electronics was certainly an important step forward. However like molecular electronics the unacceptable large contact resistance problem persisted. Besides, both patterning and device architectures are still daunting challenges.

The great promise of graphene-based electronics is that it overcomes the problems both of molecular electronics and carbon nanotube-based electronics while retaining their essential features (*1*). It specifically overcomes the patterning issues, but most importantly it obviates the contact problem. Graphene ribbons are essentially similar to carbon nanotubes exhibiting similar quantum confinement effects (*6*)(*7*) as well as high conductivity (*7*). But its most important feature is that graphitic structures can be seamlessly interconnected (Fig.1c, d) so that there is no dissipation between the functional structures (*1*). The resistance at these junctions manifests as quantum mechanical transmission and reflection coefficients (*8*), therefore the contacts themselves do not heat up, eliminating electromigration and contact failure, the fundamental weaknesses of nanoelectronic circuitry. Moreover in contrast to ordinary metal to molecule contacts, junctions between ribbons can maintain quantum mechanical phase coherence allowing in principle phase coherent quantum mechanical device structures (*6*)(*7*). Indeed graphene offers the possibility of nanoscopic interconnected structures that maintain phase coherence even at room temperature thus allowing an entirely new electronics paradigm (*1*).

These considerations were originally put forth in 2001 by de Heer and coworkers (*1*) (*9*). However the monumental task of realizing graphene electronics remained. The first major challenge was to find a platform. Among the various alternatives considered, silicon carbide was considered to be the most viable candidate (*1*). A review of the properties of epitaxially grown graphene is included in this volume (*10*). The second challenge was to develop suitable microelectronics processing methods. The third challenge is in the development of suitable dielectrics and contacting methods. Significant progress has been made and the first commercial graphene-based devices will be realized in the near future.

In this brief review we discuss progress in realizing actual graphene-based electronics concentrating primarily on transport properties.

## 2. Early Developments

Epitaxial graphene on SiC was first observed by van Bommel et al in 1975 (*11*). They noticed that when hexagonal silicon carbide was heated in ultrahigh vacuum to temperatures above 1000 C in vacuum a thin graphitic layer grew on the silicon carbide surfaces. Early interest in epitaxial graphene was focused on providing a means to control electronic contact to SiC which is an important semiconducting material. These studies were followed by many others (*12*)(*11*)(*13*)(*14*)(*15*)(*16*)(*17*)(*18*)(*19*). Monolayers were grown and identified. (*18*)(*19*) Graphene multilayers grown on the silicon-terminated face were found to be Bernal stacked (*20*)(*21*) (as in natural graphite) while graphene grows in a unique rotational phase on the carbon face (*22*) (*10*).

In 2002 and 2003 the Georgia Tech research group developed microelectronics lithography methods to pattern these epitaxial graphene initially focusing primarily on the silicon face. These results published in 2004 demonstrated the viability of graphene-based electronics (*1*). Electronic mobilities of these prototypes ($\mu \sim 1100$ $cm^2V^{-1}s^{-1}$) were greater than for typical Si-based devices, which was an important motivating factor for the field. Moreover, the two-dimensional nature of these graphene layers was clearly demonstrated, in particular by the characteristic extreme anisotropy of the magnetoresistance measurement (*1*). In addition it was shown that these graphene layers could be gated. Significantly these experiments actually presented the first example of transport in monolayer graphene (see below). It was found early on that the mobility of graphene grown on the carbon terminated face was systematically greater than graphene grown on the silicon terminated face (*7*)(*9*). Furthermore significant advances have been made in graphene growth technology (*7*)(*23*)(*24*). High mobility graphene was grown using the so-called furnace method, in which the silicon carbide chips were enclosed in a graphitic chamber and inductively heated (*7*)(*9*)(*25*)(*26*).

While the concept and early development of epitaxial graphene-based electronics (*1*) preceded exfoliated graphene physics (*27*) much of the subsequent developments occurred in parallel with little interaction. The two-dimensional electron gas community immediately embraced exfoliated graphene, because the oxidized silicon substrate provided an easy means of gating atomically thin carbon flakes (*28*). The electronics community recognized the importance of epitaxial graphene as a new platform for electronics, something not offered by exfoliated graphene. The explosive growth in graphene physics and electronics resulted from the fortuitous coincidence of these two efforts (*27*). Nevertheless from the outset, the real possibility of graphene-based electronics (*1*) was the strongest motivating factor for graphene science and technology.

Despite the overlap between these two directions they actually do represent two separate fields. Epitaxial graphene science and technology is pragmatic and not constrained to a single graphene sheet. Although epitaxial graphene is a serious candidate material for graphene-based electronics, this review primarily emphasizes its great scientific significance.

## 3. Epitaxial Graphene Structure Summary

We next summarize the morphology of graphene grown on silicon and carbon faces. A complete review can be found in this issue (*10*).

For graphene grown on the silicon face of silicon carbide, the interface terminates in a carbon rich layer called the buffer layer (*18*)(*29*)(*30*)(*31*)(*32*). While the exact atomic structure of this layer is not known its atomic density is close to that of a graphene monolayer (*33*)(*31*)(*34*). The bonding with the silicon carbide substrate is strong enough to create a bandgap so that this layer does not contribute to the transport (*30*). This layer provides isolation from bonds to the silicon carbide substrate (*35*). The graphene layer on top of the buffer layer is the first to display the characteristic graphene structure (*36*). This first graphene layer is found to be negatively charged ($n \approx 5 \cdot 10^{12}/cm^2$), while the charge density decreases rapidly in the subsequent layers with a decay length for the charge density that is somewhat larger than one layer spacing (*37*)(*38*)(*39*). However, the energy bands are slightly shifted from the charge neutrality point relative to the bands below it. This has been explained in various models (many body interactions, and a small band gap (*40*)(*41*)). The somewhat controversial bandgap issue is still under investigation(*42*)(*43*)(*44*)(*45*)(*46*)(*47*)(*48*).

The second graphene layer (*49*)(*50*) exhibits parabolic bands as observed and predicted for bilayer graphene (*21*)(*39*)(*51*). The electronic structure converges to that of bulk graphite as the number of layers is increased, consistent with their Bernal stacking (*49*)(*40*).

The structure of graphene grown on the carbon face is different from that of graphene grown on the silicon face. It is found that the first graphene layer binds tightly to the silicon carbide surface, which itself may be carbon rich possibly insulating subsequent layers from the interactions with the substrate (*50*). In contrast to silicon face graphene these layers exhibit a rotational order where alternate layers are rotated by 30° (*52*)(*10*). This structure varies depending on the growth conditions. Graphene grown in Ultra High Vacuum (UHV) conditions is rotationally disordered (*34*)(*25*)(*50*), while graphene grown using the furnace technique can show a high degree of order

(*52*). This unusual rotational stacking structure has an important consequence in that it causes the graphene layers to be electrically decoupled, as is evident from the Angular Resolved Photoemission Spectroscopy (ARPES) measurements, from transport measurements (see below), and from Raman measurements (Fig. 1e). Consequently each layer in the graphene stack is electronically similar to an independent monolayer (*22*). For this reason graphene grown on the carbon face is called multilayered epitaxial graphene (MEG). It is distinguished from Si-face multilayers, which are in fact ultrathin graphite (the literature refers to this material as "few layer graphene"). For further details see (*10*) in this issue.

For both the carbon and silicon face the topmost graphene layer is found to cover the entire surface without interruptions or breaks (*35*)(*53*). MEG is found to be extremely flat with isolated pleats (or folds) that occur about every 10 to 20 μm (Fig. 1a). These pleats are several nanometers high and result from the thermal expansion mismatch of the graphene layer and the substrate. They appear not to significantly affect the transport.

The rotational structure of MEG is revealed in scanning tunneling microscopy images that show the characteristic moiré patterns caused by the interference of the top most layer with layers below (*35*)(*53*).

### 4. Electronic Properties

The special bandstructure of graphene clearly is at the heart of its importance as a new electronic material. The linear dispersion manifested in the Dirac cone implies that the carrier velocity $v_F \approx 10^8$ cm/s is independent of its energy (*52*)(*10*). Consequently electrons in graphene ribbons resemble electromagnetic waves in a waveguide. The energy scale is approximately $E_n$ (eV) ≈ 1/W where W is the ribbon width in nm (*1*). This further implies that a bandgap of this magnitude opens in graphene nanoribbons due to quantum confinement in the ribbon (*6*)(*54*)(*55*). We next investigate the electronic properties of epitaxial graphene, which has mainly focused on MEG (i.e. multilayered graphene grown on the carbon face on 6H SiC and 4H SiC)

The electronic properties of MEG has been probed by ultrafast optical spectroscopy (*56*)(*57*), by scanning tunneling spectroscopy (*58*)(*53*)(*59*), and infrared spectroscopy(*60*)(*61*)(*62*)(*63*). The fast carrier dynamics was probed using pump probe methods (*56*). Electron hole pairs were created using a femtosecond laser pulse and the dynamics of the hot electrons was interrogated with a second laser. Relaxation times range from 1 ps for doped layers to 4 ps for undoped layers corresponding to mean free paths of the order of 1 to 4 μm(*56*). These methods were also used to determine the doping density of the layers that were similar to those found in ARPES measurements for the silicon face as evidenced from the shift in the position of the Fermi level with increased graphene thickness (*52*). Decay lengths of the charge density corresponded to about one monolayer (*64*)(*56*) and the doping density of the interface is of the order of 5 $10^{12}$/cm$^2$ corresponding to a Fermi level that is approximately 300 meV above the Dirac point.

Recent scanning tunneling spectroscopy measurements in high magnetic fields have yielded new insight into the graphene Landau levels (*53*). Experiments performed at NIST probed the Landau levels of the topmost graphene layers in multilayered epitaxial graphene (Fig.2). These experiments directly demonstrate that the graphene layers in MEG are indeed decoupled and that at least the topmost layer is electronically similar to an isolated graphene layer. They further provide a new perspective on the quantum Hall state in graphene since the Landau levels can be probed locally as a function of position on the graphene layer. Indeed these spectra reveal interesting fine structure (eg. the n=0 level is split). The fine structure features of the Landau levels are currently under investigation.

Potemski and coworkers have exhaustively studied the infrared absorption properties of MEG (Fig. 3) (*60*)(*61*)(*62*)(*63*). Their studies conducted in magnetic fields up to 32 Tesla demonstrated for the first time the characteristic square root magnetic field dependence of the energy $E_n$ of graphene Landau levels: $E_n = \pm c\sqrt{2e\hbar B|n|}$, as they are for an ideal graphene layer (*60*). They further demonstrated that most of the graphene layers in multi-layered epitaxial graphene were electronically decoupled. Further studies demonstrated insignificant broadening of the Landau levels with increasing temperature indicating that the electron-phonon coupling in these C-face layers is weak (*60*). The weak electron-phonon coupling is also reflected in the temperature dependence in early transport measurements (*7*). Landau levels were observed at magnetic fields as low as 40 mT at 4 Kelvin and below 1 T at room temperature (*60*). The measurements show that the charge density of the undoped layers is less than $5\ 10^9/cm^2$ and that the Landau level lifetime is of the order of 0.1 ps. These results indicate that Landau levels may be used in magnetic devices (on a 100 nm length scale) operating at room temperature.

## 5. Electronic Transport

We next examine the electronic transport properties of epitaxial graphene on both faces of hexagonal silicone carbide. Currently those faces are candidates for graphene-based electronics and both faces have interesting fundamental 2DEG properties.

*Transport in Si-face epitaxial graphene.*

As mentioned before, the earliest graphene transport measurements were performed on the Hall bars patterned on 6H SiC, Si-face samples using the UHV production method.(*1*) The material was of relatively poor quality that was reflected in the low mobilities in most of the samples. Square resistances ranged from one to several hundred kΩ per square. Nevertheless these samples clearly indicated the two-dimensional electron gas properties of epitaxial graphene and most of the important transport features of epitaxial graphene. We reproduce here the Hall measurements published in 2004 (Fig. 4a). Note that the properties of sample "A" corresponds to those of a single graphene sheet. This assessment is based on the fact that the charge density derived from the Hall slope corresponds very well to that found from the SdH oscillations (as it should). This is not the case for multilayered samples (*9, 75*) where significant decreases of the Hall coefficient are observed that are not reflected in the SdH oscillation positions. Furthermore, the graphene thickness measurements at the time relied on Auger electron spectroscopy that significantly overestimated the thickness of the graphene layer by not taking the buffer layer into account. The magnetic field strengths were limited to 8 Tesla. The Shubnikov the Hass (SdH) oscillations are clearly evident and correspond to the 2$^{rd}$ and 3$^{rd}$ Landau levels. The graphene layer had a mobility of 1100 cm² V$^{-1}$s$^{-1}$ at 4K. For comparison, Fig. 4 also shows later measurements on a silicon face graphene monolayer Hall bar (from ref (*65*)). While not realized at the time (since the inertness of the buffer layer was not yet known) these measurements are in fact the first transport measurements of monolayer graphene (*28*).

Since then several groups have performed transport measurements on 6H silicon face epitaxial graphene monolayers (*66*)(*65*)(*67*). Mobilities are found to reduce significantly with increasing temperature. Typical low-temperature mobilities are about 2000 cm² V$^{-1}$ s$^{-1}$. Mobilities degrade significantly with temperature and typically fall below 1000 cm² V$^{-1}$ s$^{-1}$ at room temperature (*66*). The half-integer quantum Hall effect has recently been observed by several groups in monolayer silicon face epitaxial graphene (*66*)(*65*)(*67*). Even though the mobilities are relatively small, the half-integer quantum Hall effect is well-established and it has been proposed as a resistance standard (*67*).

*Transport in C-face graphene: MEG*

The vacuum furnace graphene production technique significantly improved the quality of the graphene layers. Regardless, graphene produced on the silicon terminated face typically did not exceed mobilities of 2000 cm² V$^{-1}$s$^{-1}$. On the other hand mobilities of the graphene layers produced on the carbon face often exceeded 10,000 cm² V$^{-1}$s$^{-1}$ and do not significantly depend on temperature (Fig. 4b inset) (*7*)(*9*), consistent with the IR measurements (*60*). However, in contrast to the silicon terminated face, it is more difficult to control the thickness of the graphene layer on the carbon terminated face. The clearly superior transport properties warranted the efforts to improve this control, and currently C face monolayer graphene can be grown routinely (*68*).

However, MEG transport is also of considerable interest. Figure 4b shows the measurements on 500 nm wide MEG ribbon composed of 10 graphene layers (*7*) Analysis of the Shubnikov-de Haas oscillations Fig. 5c (see also Figs. 5a,b for a 1.5µm wide Hall bar up to 24 T) show that the transport layer at the interface is electronically similar to monolayer graphene with a Berry phase of π indicating that it's electronically decoupled from the layer above it. (As mentioned before it is now known that the electronic decoupling results from the special rotational stacking in epitaxial graphene (*22*)). The charge density of this layer is approximately $5 \times 10^{12}$ cm$^{-2}$. The low temperature mobility was 27,000 cm$^2$V$^{-1}$s$^{-1}$. Quantum confinement effects were also evident from the low magnetic field response (Fig. 5c).

The Georgia Tech group has persistently pursued MEG research (*7*)(*69*)(*9*)(*25*)(*22*)(*70*)(*71*). While it is often believed that science and technology require monolayer graphene, this prejudice is unfounded. In fact multilayers have distinct advantages: a multilayered ribbon can be expected be more defect tolerant than a monolayer. Moreover, its noise figures are expected to be superior to that of a monolayer (*72*). Furthermore the topmost layer can be chemically converted to a semiconducting (or insulating) form of graphene by chemical passivation (*73*)(*74*). Such a layer may be used either as a dielectric or to support a dielectric on top of it, with minimal disruption of the transport layer under it.

The 2DEG properties of MEG have been investigated using standard Hall bar structures. It was found that the SdH oscillations were essentially quenched (*9*) (Fig.6). This unexpected result suggested that in these 2-D samples additional dissipation was induced by the uncharged layer on top of the transport layer. In fact calculations by Darancet et al (*75*) showed that this was indeed the case. They found a magnetic field dependent coupling between the layers, which explained the quenched magnetoresistance oscillations. Note that the mechanism requires two (quasi-infinite) 2-D graphene sheets, of which one is charged and the other is not, as is the case of a multilayered epitaxial graphene. This explanation has further merit because it explains why the SdH oscillations are prominent in graphene ribbons: in that case the neutral overlayer does not have the required electronic structure to quench the magnetoresistance oscillations of the transport layer. Nevertheless the high degree of structural purity in graphene may itself not support magnetoresistance oscillations, which (like the quantum Hall effect) require some disorder.

One important property that distinguishes graphene from a normal 2DEG is that it exhibits weak anti-localization as demonstrated by Wu et al. (*76*) (Fig. 5d-e). This property manifests as an increase in the resistance with increasing magnetic field. It results from the pseudo-spin character of the carriers in graphene and has a characteristic temperature and field dependence (*77*). The effect is typically not observed in exfoliated graphene samples because of surface roughness induced disorder. (*78*)

Recently monolayer graphene has also been produced on the C-face of 4H SiC (*68*). The transport properties of this layer have been interrogated in standard Hall bar structures. Figure 4d shows magnetotransport measurements in a Hall bar patterned over a series of silicon carbide steps. The sample exhibits the quantum Hall effect at low temperatures. The SdH oscillations in low fields give way to the characteristic zero longitudinal resistance regions in high fields corresponding to quantum Hall plateaus in the transverse resistance. It is important to note that although the Hall bar is draped over the substrate steps (and is significantly contaminated) it has both a higher mobility and a better developed quantum Hall structure than a Hall bar patterned on a single terrace (*68*), which is relevant for its application potential. The latter shows only extremely weak magnetoresistance oscillations reminiscent of the oscillations observed in a multilayer epitaxial graphene. Nevertheless the latter is certainly a monolayer as evident from the linear Hall effect as well as from ellipsometry measurements.

## 6. Chemical Modification And Functionalization

Graphene can be chemically modified. This has been demonstrated in several experiments in which the graphene surface is functionalized with various molecules (Fig. 7) (*74*)(*73*)(*79*). The significance of chemical modification is that it can convert the graphitic $sp^2$ bonds to diamond like $sp^3$ bonds (loosely speaking, graphitic carbon is transformed into diamond like carbon). This chemical conversion will therefore produce significant band gaps in the electronic structure of the graphene layer. Two experiments on epitaxial graphene have demonstrated this effect (Fig. 7d). In one case epitaxial graphene was converted locally to graphene oxide using the Hummer's method and the bandgap was demonstrated by its Schottky barrier(*79*). In the other case the surface of a multilayered architecture graphene sample was functionalized with Aryl molecules (Fig. 7a-b) (*74*)(*73*). The bandgap was detected in ARPES measurements, which further attested to the high quality of the functionalized graphene(*10*). These experiments have provided proof of principle evidence of the effectiveness of chemical modification. Further tests are required to determine if the mobility of the functionalized material is sufficient for applications. On the other hand the functionalized surface graphene layer may be used as a dielectric to gate the layer below it. There is no doubt that chemistry will play an important role in the development of graphene-based electronics.

## 7. Devices

Before devices can become commercially viable, epitaxial graphene must be fully developed as a new electronic material. The previous discussion shows that great advances have been made in our understanding of the material and the first commercial epitaxial graphene devices are on the horizon. Above we discussed the quantum Hall effect resistance standard, and here we briefly review the status quo of conventional electronic devices.

Figure 8 shows three examples of epitaxial graphene transistor prototypes (*70*). It shows two top gated transistors (Fig. 8a-d) using the conventional geometry consisting of a graphene channel coated with a dielectric and metal gate. This third structure consists of a side-gated transistor (Fig. 8e). This is an all graphene device where a narrow graphene channel is flanked by graphene side gates. Figure 9 shows the first example of several dozen epitaxial graphene Field Effect Transistors (FET) patterned on a single epitaxial graphene chip (both on the C-face and on the Si-face) (*80*). These first results show that the epitaxial graphene electronics is on the right track for a new carbon-based electronics.

It is clear that the characteristics of the transistors mentioned above are far from ideal and that they cannot compare with CMOS FETs. The primary reason is that 2-D graphene does not have a bandgap, so that the field effect is dominated by the modulation of the charge density by the gate potentials. Moreover in contrast to normal 2DEGs, the mobility is found to be approximately inversely proportional to the charge density, and the conductivity depends (approximately) linearly on the applied field with a minimum conductivity of about 0.1 mS. Ultimately, the FET on-to-off resistance ratios are of the order of 10, which is minute compared to CMOS FETs ratios that exceed $10^7$.

Enhancing the on-to-off resistance ratios will require a bandgap. There are two ways to accomplish this. One is to manufacture graphene ribbons that are narrower than 10 nm (to produce a bandgap >100 meV), and the other is to chemically convert the graphene to a semiconducting form with a bandgap. The former method will require lithography methods that are potentially damaging and the effect on the mobility of the ribbons is still not determined. The latter method does not require state of the art lithography, but the chemical conversion may also significantly affect the mobility. In any case, it is clear that graphene electronics is currently not poised to replace CMOS.

However graphene-based electronics is not expected to compete with silicon-based electronics but rather to complement it. For example, graphene can outperform silicon-based electronics in speed and certain high-speed devices do not necessarily require large on-to-off resistance ratios.

Significant advances have been made in demonstrating the high-speed capabilities of graphene-based electronics. Recent results reported from the HRL laboratories (*81*) (Fig. 10) and IBM (*82*) have shown analog monolayer Si-face epitaxial graphene FET's with larger than unity gain at operating speeds in excess of 10 GHz. Moreover speeds in excess of one THz having are predicted to be feasible. Note that the silicon carbide substrate has a distinct advantage over, for example, a silicon oxide substrate. The optical phonon frequencies, that limit the mobility especially at high temperatures are high, 115 to120 meV, which is about twice as high as for silicon dioxide.

Higher-speed devices will require even higher mobility material as provided for example in monolayer C face epitaxial graphene (where room temperature Hall mobilities > 10,000 $cm^2 V^{-1} s^{-1}$) are measured. Note that MEG graphene layers have been measured with mobilities exceeding 200,000 $cm^2 V^{-1} s^{-1}$ at room temperature (*60*).

## 8. Conclusion And Outlook

Epitaxial graphene is a new material that is revolutionizing and revitalizing low dimensional electron gas physics. While conventional semiconductor electron gas physics and exfoliated graphene flake physics is essentially limited to transport measurements, the high quality of epitaxial graphene also allows interrogation with a variety of surface sensitive probes, optical probes and advanced light source structural probes, yielding impressive new science.

While the field of epitaxial graphene electronics is still in its infancy, the progress in the past few years has been remarkable. Epitaxial graphene on both the C- and the Si-face has shown its applications potential. The first devices may be on the market soon and they will probably consist of analog high-speed epitaxial graphene transistors for specialized applications and possibly epitaxial graphene quantum Hall effect resistance standards. However as we are learning more and more about this remarkable new material it will probably become clear that new electronic paradigms (like coherent electronics and spintronics) are feasible. We already see a remarkable difference between the evolution of epitaxial graphene electronics and its predecessor, carbon nanotube-based electronics. The latter produced remarkable prototypes, however solutions to the daunting technological problems involving lithography and contacts were never found. In contrast, in epitaxial graphene electronics most of those problems were solved at the outset.

It is not reasonable to expect that graphene-based electronics will replace silicon-based electronics. The two undoubtedly will ultimately evolve in parallel much like aviation and navigation developed in parallel, both essentially accomplishing the same task but in different ways and for different purposes.

The remarkable scientific opportunities offered by multilayered epitaxial graphene cannot be overestimated. Currently, graphene research has singled out "isolated" graphene monolayers as the most important players in the field, and this point of view is clearly influenced by the current theoretical tractability of the monolayer compared with multilayers rather than its intrinsic greater scientific value. The unique properties of the multilayer epitaxial graphene may bring this material to the forefront of graphene research.

**Acknowledgements**
This research was supported by the W.M. Keck Foundation, the Partner University Fund from the Embassy of France and the NSF under Grant No. DMR-0820382.

Figure Captions

Figure 1

Typical AFM images of furnace grown graphene layers on a C-face (a) and Si-face (b). (a): the C-face multilayered epitaxial graphene layer is atomically flat over tens of microns. The layers are continuous, and draping over substrate steps without breaks. The white lines are pleats in the graphene sheet that do not disrupt the continuity of the layer. (b) several graphene layers are draped over the SiC step structure (c): SEM picture of patterned Hall bar structure. The ribbon is patterned on a single terrace, with graphene pads extending out towards the Pd/Au contacts. (d) Example of integrated structures on a SiC chip, featuring a pattern of evaporated gold pads connecting a hundred micron and sub-micron size graphene ribbons (not seen) grown on the C-face of 4H-SiC. The background contrast is an artifact from the tape on the back of the transparent SiC chip. (e) Raman spectrum of multilayered epitaxial graphene. The SiC background has been subtracted. (Inset) The 2D peak can be fitted with a single Lorenzian ( 25 cm$^{-1}$ width) indicating that the layers are decoupled. In particular the shoulder at low energy for Bernal staked graphite in not observed. The disorder induced D peak at 1350 cm$^{-1}$ is not observed, indicating that the extended graphene layers have a very low density of defects.

Figure 2

Electronic structure of MEG (top layer) from scanning tunneling spectroscopy (STS) performed in a magnetic field. (a) Foreground shows a cartoon of the quantized cyclotron orbits (Landau levels) probed by STS. In the background is an STM topograph of the sample showing the graphene atomic honeycomb and a small (≈0.01 nm) modulation in the apparent height due to the moiré alignment of layers. (b) Inset: Landau level energy structure. Landau levels lie at discrete cyclotron energies. The data shows tunneling magnetoconductance oscillations (TMCO) detected in the tunneling conductance dI/dV. A peak in the dI/dV results when a Landau level coincides with the bias voltage of the sample $V_B$. (c) At a fixed magnetic field, the LLs appear as peaks in the dI/dV as the sample bias is changed ($B$=5 T for this spectrum). The inset shows that the Landau energies correspond to those of single-layer graphene. (d) Both the TMCO measurements of (b) and the conventional STS in (c) imply a linear $E(k)$ relation. Shown here are the TMCO energy bands. Figure adapted from reference (*53*).

Figure 3

Infra-red spectroscopy demonstrating the graphene Landau level structure of MEG on the C-face. (a) the relative transmission of light at very low field. The absorption line corresponds to an inter-Landau-level transition. (b) The plot of the field dependence of the absorption energy demonstrates that the Landau levels separation disperses as √B as expected for graphene. Several transition lines are observed, as indicated for the first six lines. The inset is a schematic of the allowed transitions (same colors); as the field increases the high index levels start to depopulate, as indicated by the vertical black lines, and more transitions are allowed, until the Fermi level (dotted line) intersects the last level (n=1). The minimum field at which the n=0→n=1 (labeled $L_{0(-1)}$ →$L_{1(0)}$ - grey) transition is observed (40mT) is therefore a measure of the Fermi level $E_F$=8meV from the Dirac point ($n_s$=5x10$^9$/cm$^2$). (c) The $L_{0(-1)}$ →$L_{1(0)}$ absorption peak at low field as a function of temperature and (d) a plot of the peak position, width and area with temperature. The peak area decreases as the Boltzmann occupation factor of the level. The peak width is

*independent* of temperature indicating very weak electron-phonon scattering process and large scattering time (τ>100 fs, comparable with the weak anti-localization of Fig 6), yielding a room temperature mobility μ>250,000 cm$^2$/Vs. (Figure adapted from Ref. (*60*))

Figure 4

2D transport $\rho_{xx}$ and $\rho_{xy}$ measured for epitaxial graphene Hall bars. (a) first transport measurement of monolayer graphene, measured on a patterned Hall bar on the Si face of 6H SiC from Refs. (*1*) and (*9*). (400 μm X 600 μm, mobility μ = 1200 cm$^2$/V·s, coherence length $l_\Phi$≈300 nm at T=4K). The magnetoresistance $\rho_{xx}$ at T =0.3, 2 and 4 K (resp. black, blue, grey curves) shows well-developed SdH peaks, indicated with their Landau indices n; $\rho_{xy}$ at 0.3 K (red), shows a weak feature at the expected Hall plateau position. The amplitude of the weak localization peak at B = 0 corresponds to 1$G_0$. (b) MEG on the C-face (0.5μm x 6μm, μ=27,000 cm$^2$/Vs, $l_\Phi$≈1.1 μm at T=4K from Ref. (*7*)) for temperatures ranging from 4K to 58K. The SdH oscillations correspond to Landau indexes 4 to 25. The Landau plot shows that the Berry phase is π (see also Fig 6), as it is for a graphene monolayer, showing for the first time the electronic decoupling of the layers in MEG. Inset: temperature dependence of the resistivity for the same sample. (c) from Ref (*65*). Half integer quantum Hall effect in monolayer epitaxial graphene on the Si-face samples (10μm X 30μm, μ=3,600 cm$^2$/Vs) measured at 0.8K up to 18T. The charge density is comparable to Fig 4a; note the similarity of the resistivity components $\rho_{xx}$ and $\rho_{xy}$ with Fig.4a (*1*)(*9*). At high field, the Hall resistance shows characteristic Hall plateaus at $\rho_{xy}$=(h/4e$^2$) / (n+1/2), where *n* is the Landau level index, and the magnetoresistivity $\rho_{xx}$ shows characteristic SdH oscillations; the resistance vanishes for low Landau indexes consistent with the quantum Hall effect. (d) Half integer quantum Hall effect in monolayer epitaxial graphene on the C-face samples (1.8μm X 4.6μm, μ=20,000 cm$^2$/Vs) measured at 4 K and 200K. The characteristic half integer quantum Hall plateaus (red), the magnetoresistivity (black) SdH oscillations and zero resistance are clearly observed. The n=0 Hall plateau extends over more that 14 Tesla, and it is very clearly developed at high temperature up to 200K (dotted red). Inset: AFM image of the Hall bar patterned over several SiC steps (the graphene Hall bar structure is highlighted on the SiC step structure). The image shows (white spots, covering about 17% of the surface) and pleats in the graphene (white lines). The Hall bar retains its high mobility and the QHE is observed despite important contamination.

Figure 5

The individual graphene sheet in multilayer epitaxial graphene (MEG) are electronically decoupled as is clear from transport measurments. (a) The magnetoresistance presents a single SdH period up to 23 Tesla (Hall bar 1.5μm X 6.5μm, $n_s$=3.7 10$^{12}$/cm$^2$) indicating that the highly charged layer at the interface dominates transport. (b) The Landau plot (Landau index as a function of inverse magnetic field) intersects the origin, consistent with a Berry's phase of π. (c) Lifschitz-Kosevitch analysis of the temperature dependence of the SdH oscillations $A_n(T)$=u /sinh(u) where u=2π $k_B^2$/ΔE(B). The Landau level dispersion ΔE(B) is reproduced as expected from theory (open circles), and shows the low field saturation that is expected due to quantum confinement (see Ref. (*7*)). This occurs when the cyclotron diameter is larger than the ribbon width. (d) and (e) Demonstration of the weak anti-localization effect, from Ref (*76*). (d): magnetoresistance (1.4K, 4.2K, 7K, 10K, 15K, 20K, 30K) and fit using weak-antilocalization theory for graphene (Hall bar 100μm X 1000μm, μ=11,600 cm$^2$ V$^{-1}$s$^{-1}$, transport time τ~260 fs). All the curves are fitted with only one temperature dependent parameter, the phase coherence time $\tau_\Phi$=C/T, with C=20ps.K, attributed to electron-electron scattering. The weak antilocalization is dominated by valley symmetry conserving processes, consistent with scattering from long-range potentials arising from charges in the substrate. (e) comparison of the fit for weak antilocalization (solid line) and weak-localization (dashed line).

Figure 6

(a) Magnetoresistance $\rho_{xx}$ and (b) Hall resistance $\rho_{xy}$ on three patterned Hall bars on multilayered epitaxial graphene, showing the influence of the ribbon width on the amplitude of the SdH oscillations. The three ribbons have comparable resistivities, mobilities and charge densities, as determined from the period of the SdH oscillations (black: 100μm X 1000μm, μ=11,600 cm$^2$/Vs, $n_s$=4.6x10$^{12}$/cm$^2$; blue: 1μm X 5μm, μ=12,500 cm$^2$/Vs, $n_s$=45.1x10$^{12}$/cm$^2$; red: 0.5μm X 6μm, μ=27,000 cm$^2$/Vs, $n_s$=3.4x10$^{12}$/cm$^2$). (c) – (e) relative amplitude of the SdH oscillations shown in (a) for the three Hall bars as indicated, showing that the amplitude increases by two orders of magnitude as the ribbon width decreases.

Figure 7

Functionalization of graphene. (Top) Schematic of graphene functionalization by covalent attachment of aryl groups to the basal carbon atoms (after Ref. (*73*)). AFM images of epitaxial graphene before (a) and after (b) functionalization. (c) Kekulé and Clar sextet representations of functionalized graphene at 25% coverage. The chemical approach to the generalization of electronic devices in graphene allows the creation of insulating, semi-metallic and metallic regions. (d) I-V characterisitics of graphene oxide (after Ref. (*79*)). The grafting of hydroxyl and epoxy groups turns graphene into a semiconductor or insulator (blue). Partial reduction by heat treatment (red) restores conductivity, allowing a tuning of the transport properties.

Figure 8

Epitaxial graphene field effect transistors (after Ref. (*70*)) (a) Conductivity σ as a function of gate voltage at 300K for a single graphene layer on Si-face SiC (Hall bar 3.5 μm x 12.5μm). The ratio of maximum to minimum resistance is $I_{on}/I_{offf}$=31. The minimum conductivity is close to 2e$^2$/πh as for exfoliated graphene flakes (dotted line). Inset: the measured top gated ribbon, before and after gate deposition (spin-on HSQ resist and evaporated metal gates on top). (b) Resistivity (black) and Hall resistance (red) as a function of gate voltage at 5 Tesla and 300K. The resistivity peaks when $\rho_{xy}$ changes sign. Inset: temperature dependence of $\rho_{xx.}$ (c) and (d) Conductivity as a function of gate voltage for a C-face multilayer epitaxial graphene Hall bar, and the device picture. Three gates (G1, G2 and G3, light color) evaporated on top of the dielectric (light brown rectangle) cover partially the ribbon laying between the current leads I. (c): conductivity for a portion of the ribbon entirely covered by the gate (gate G1, voltage probe V1 and V2). (d) partially gated ribbon ((1) gate G1, (2) gate G2). Depending on the conditions, the gated portion of the ribbon can be p- (1) or n-doped (2). (e) Conductivity for a C-face multilayer epitaxial graphene nano-ribbon (width 50 nm) with split gates, and SEM image of the gated structure.

Figure 9

Array of patterned top gated epitaxial graphene FETs. 100 FET transistors have been patterned on 3 X 4mm$^2$ SiC chip (channel 5μm X 10μm, Al gate on top of 40nm evaporated HfO$_2$). (a) Optical image of the transistors; the scale bar is 100μm. (b) $I_{sd}/V_{sd}$ vs $V_g$ is reproduced on the various Si-face FETs. The minimum at $V_g$=-1 indicates a charge density $n_s$=3.8x10$^{12}$/cm$^2$, consistent with gating of the charged interface layer. A large on-off ratio is observed (>10). (c) multilayer EG on the C-face. The minimum close to zero indicated gating of the top – quasi neutral layers. The gated top layer is partly shorted by the charged interface layer that is not gated in this multilayered sample. Note that the gate fields have a screening length of about 1 layer). (After Ref. (*80*))

Figure 10

Ultrahigh frequency graphene FETs produced on a 50mm graphene wafer processed using standard lithographic techniques. Measured magnitude of extrinsic and unilateral gain are shown as a function of frequencies with Vds = 5 V and Vgs = −2.5 V. The current gain cutoff frequency of this device is 4.4 GHz (after Ref. (*81*)). Inset (a): processed graphene wafer. Inset (b): scanning electron micrograph of monolayer Si face graphene FET with a 2 μm x 12 μm channel.

1. C. Berger et al., *Journal of Physical Chemistry B* **108**, 19912 (2004).

2. W. Lu, C. Lieber, *Nature Materials* **6**, 841 (2007).

3. J. Heath, *Annual Review Of Materials Research* **39**, 1 (2009).

4. R. P. Avouris, *Accounts of Chemical Research* **35**, 1026 (2002).

5. S. Frank, P. Poncharal, Z. L. Wang, W. A. Heer, *Science* **280**, 1744 (1998).

6. K. Nakada, M. Fujita, G. Dresselhaus, M. S. Dresselhaus, *Physical Review B: Condensed Matter and Materials Physics* **54**, 17954 (1996).

7. C. Berger et al., *Science* **312**, 1191 (2006).

8. K. Wakabayashi, M. Sigrist, *Physica B* **284**, 1750 (2000).

9. W. A. de Heer et al., *Solid State Communications* **143**, 92 (2007).

10. Sprinkle, M., *in print, J.Phys. D* (2010).

11. A. J. van Bommel, J. E. Crombeen, A. van Tooren, *Surface Science* **48**, 463 (1975).

12. I. Forbeaux, J. M. Themlin, J. M. Debever, *Physical Review B: Condensed Matter and Materials Physics* **58**, 16396 (1998).

13. C. S. Chang, N. J. Zheng, I. S. T. Tsong, Y. C. Wang, R. F. Davis, *Journal of Vacuum Science & Technology B* **9**, 681 (1991).

14. L. Li, I. S. T. Tsong, *Surface Science* **364**, 54-60 (1996).

15. L. I. Johansson, F. Owman, P. Martensson, *Physical Review B: Condensed Matter and Materials Physics* **53**, 13793 (1996).

16. P. Martensson, F. Owman, L. I. Johansson, *Physica Status Solidi B* **202**, 501 (1997).

17. F. Owman, L. I. Johansson, P. Martensson, *Silicon carbide and related materials 1995* **142**, 477 (1996).

18. A. Charrier et al., *Journal of Applied Physics* **92**, 2479 (2002).

19. I. Forbeaux, J. M. Themlin, A. Charrier, F. Thibaudau, J. M. Debever, *Applied Surface Science* **162**, 406 (2000).

20. E. Rollings et al., *Journal of Physics and Chemistry of Solids* **67**, 2172 (2006).

21. T. Ohta, A. Bostwick, T. Seyller, K. Horn, E. Rotenberg, *Science* **313**, 951 (2006).

22. J. Hass et al., *Physical Review Letters* **100**, 125504 (2008).

23. K. Emtsev et al., *Nature Materials* **8**, 203 (2009).

24. C. Virojanadara et al., *Physical Review B: Condensed Matter and Materials Physics* **78**, 245403 (2008).

25. J. Hass et al., *Applied Physics Letters* **89**, 1431 (2006).

26. J. Hass, W. A. de Heer, E. H. Conrad, *Journal of Physics: Condensed Matter* **20**, 323202 (2008).

27. A. K. Geim, K. S. Novoselov, *Nature Materials* **6**, 183 (2007).

28. K. S. Novoselov et al., *Science* **306**, 666 (2004).

29. K. Emtsev et al., in *6th European Conference on Silicon Carbide and Related Materials, ECSCRM 2006, Newcastle upon Tyne, UK, September 3rd - 7th, 2006* , (2006).

30. G. M. Rutter et al., *Physical Review B: Condensed Matter and Materials Physics* **76** , 235416 (2007).

31. K. V. Emtsev, F. Speck, T. Seyller, L. Ley, J. D. Riley, *Physical Review B: Condensed Matter and Materials Physics* **77**, 155303 (2008).

32. J. B. Hannon, R. M. Tromp, *Physical Review B* **77**, 241404 (2008).

33. R. Tromp, J. Hannon, *Physical Review Letters* **102**, 106104 (2009).

34. J. Hass, J. E. Millan-Otoya, P. N. First, E. H. Conrad, *Physical Review B* **78**, 205424 (2008).

35. J. Hass et al., *Physical Review B* **75**, 205424 (2007).

36. T. Ohta et al., *Physical Review Letters* **98**, 206802 (2007).

37. E. Rolling et al., *Journal of Physics and Chemistry of Solids* **67**, 2172 (2006).

38. D. Sun et al., *Physical Review Letters* in print (2010).

39. A. Bostwick et al., *New Journal of Physics* **9**, 385 (2007).

40. S. Y. Zhou et al., *Nature Materials* **6**, 770 (2007).

41. L. Vitali et al., *Surface Science* **602**, L127 (2008).

42. E. Rotenberg et al., *Nature Materials* **7**, 258 (2008).

43. M. Mucha-Kruczynski et al., *Physical Review B* **77**, 195403 (2008).

44. M. Polini, A. Tomadin, R. Asgari, A. H. MacDonald, *Physical Review B* **78**, 115426 (2008).

45. S. Kim, J. Ihm, H. J. Choi, Y. W. Son, *Physical Review Letters* **100**, 176802 (2008).

46. F. Varchon, P. Mallet, J. Veuillen, L. Magaud, *Physical Review B* **77**, 235412 (2008).

47. M. Polini, R. Asgari, Y. Barlas, T. Pereg-Barnea, A. MacDonald, *Solid State Communications* **143**, 58 (2007).

48. C. Park, F. Giustino, C. D. Spataru, M. L. Cohen, S. G. Louie, *Nano Letters* **9**, 4234 (2009).

49. G. M. Rutter, J. N. Crain, N. P. Guisinger, P. N. First, J. A. Stroscio, *Journal of Vacuum Science and Technology A* **26**, 938 (2008).

50. P. Mallet et al., *Physical Review B* **76**, 041403 (2007).

51. C. Riedl, U. Starke, J. Bernhardt, M. Franke, K. Heinz, *Physical Review B* **76**, 245406 (2007).

52. M. Sprinkle et al., *Phys. Rev. Lett.* **103**, 226803 (2009).

53. D. L. Miller et al., *Science* **324**, 924-927 (2009).

54. M. Y. Han, B. Ozyilmaz, Y. Zhang, P. Kim, *Physical Review Letters* **98**, 206805 (2007).

55. X. Wang et al., *Physical Review Letters* **100**, 206803 (2008).

56. D. Sun et al., *Physical Review Letters* **101**, 157402 (2008).

57. P. Plochocka et al., *Physical Review B* **80**, 245415 (2009).

58. G. M. Rutter et al., *Science* **317**, 219 (2007).

59. N. P. Guisinger, G. M. Rutter, J. N. Crain, P. N. First, J. A. Stroscio, *Nano Letters* **9**, 1462 (2009).

60. M. Orlita et al., *Physical Review Letters* **101**, 26760 (2008).

61. M. L. Sadowski, G. Martinez, M. Potemski, C. Berger, W. A. de Heer, *Physical Review Letters* **97**, 266405 (2006).

62. M. L. Sadowski, G. Martinez, M. Potemski, C. Berger, W. A. De Heer, *International Journal of Modern Physics B* **21**, 1145 (2007).

63. P. Plochocka et al., *Physical Review Letters* **100** , 087401 (2008).

64. P. Lauffer et al., *Physical Review B: Condensed Matter and Materials Physics* **77**, 155426 (2008).

65. T. Shen et al., *Appl. Phys. Lett.* **95**, 172105 (2009).

66. J. Jobst et al., *ArXiv :*0908**.**1900 (2009).

67. A. Tzalenchuk et al., *Nat Nano* **5**, 186 (2010).

68. X. Wu et al., *Applied Physics Letters* **95**, 223108 (2009).

69. C. Berger et al., *Physica Status Solidi (a)* **204**, 1746 (2007).

70. X. Li et al., *Physica Status Solidi A* **207**, 286 (2010).

71. M. Sprinkle, P. Soukiassian, W. A. de Heer, C. Berger, E. H. Conrad, *Physica Status Solidi (RRL) - Rapid Research Letters* **3**, A91-A94 (2009).

72. Y. Lin, P. Avouris, *Nano Letters* **8**, 2119 (2008).

73. E. Bekyarova et al., *Journal of the American Chemical Society* **131**, 1336 (2009).

74. E. Bekyarova, M. E. Itkis, P. Ramesh, R. C. Haddon, *physica status solidi (RRL)* **3**, 184 (2009).

75. P. Darancet, N. Wipf, C. Berger, W. A. de Heer, D. Mayou, *Physical Review Letters* **101**, 116806 (2008).

76. X. Wu, X. Li, Z. Song, C. Berger, W. A. de Heer, *Physical Review Letters* **98**, 136801 (2007).

77. E. McCann et al., *Physical Review Letters* **97**, 146805 (2006).

78. S. V. Morozov et al., *Physical Review Letters* **97**, 016801 (2006).

79. X. Wu et al., *Physical Review Letters* **101**, 026801 (2008).

80. J. Kedzierski et al., *IEEE Transactions on Electron Devices* **55**, 2078 (2008).

81. J. Moon et al., *Electron Device Letters, IEEE* **30**, 650-652 (2009).

82. Y.-M. Lin et al.*, Science* **327**, 662 (2010).

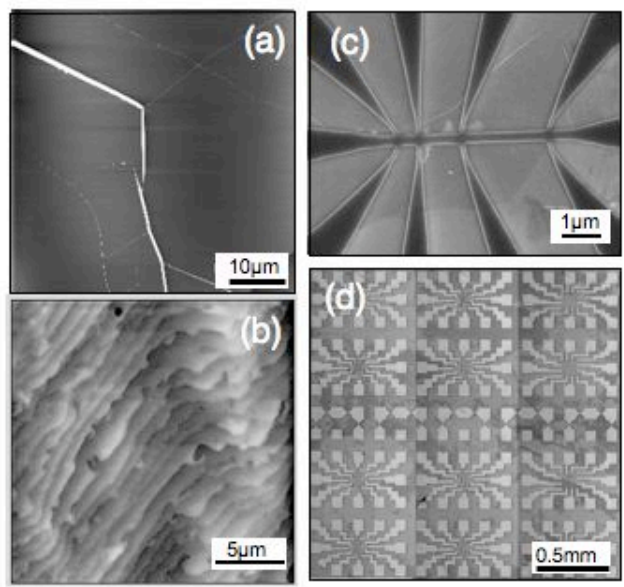
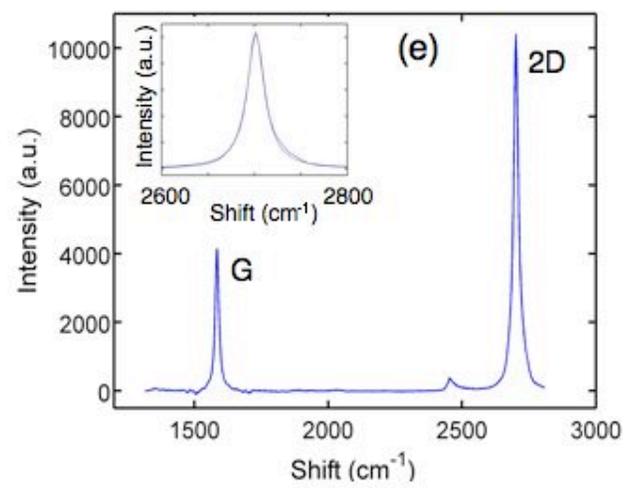

Figure 1

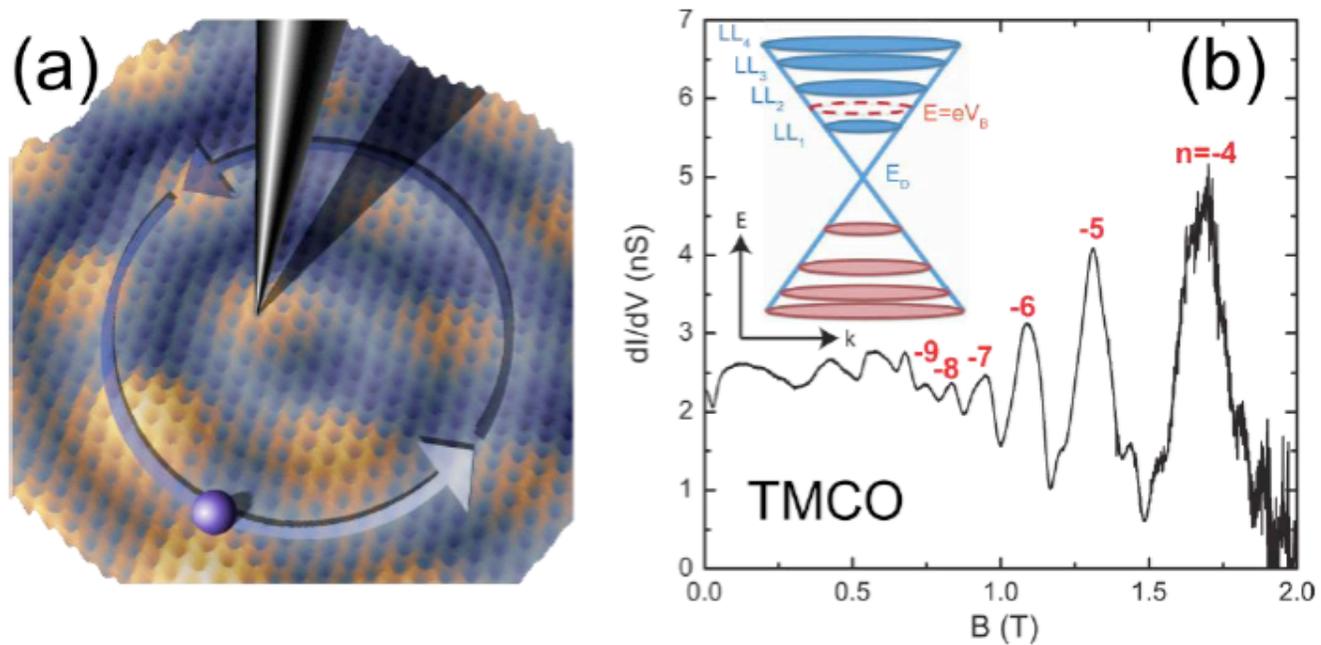
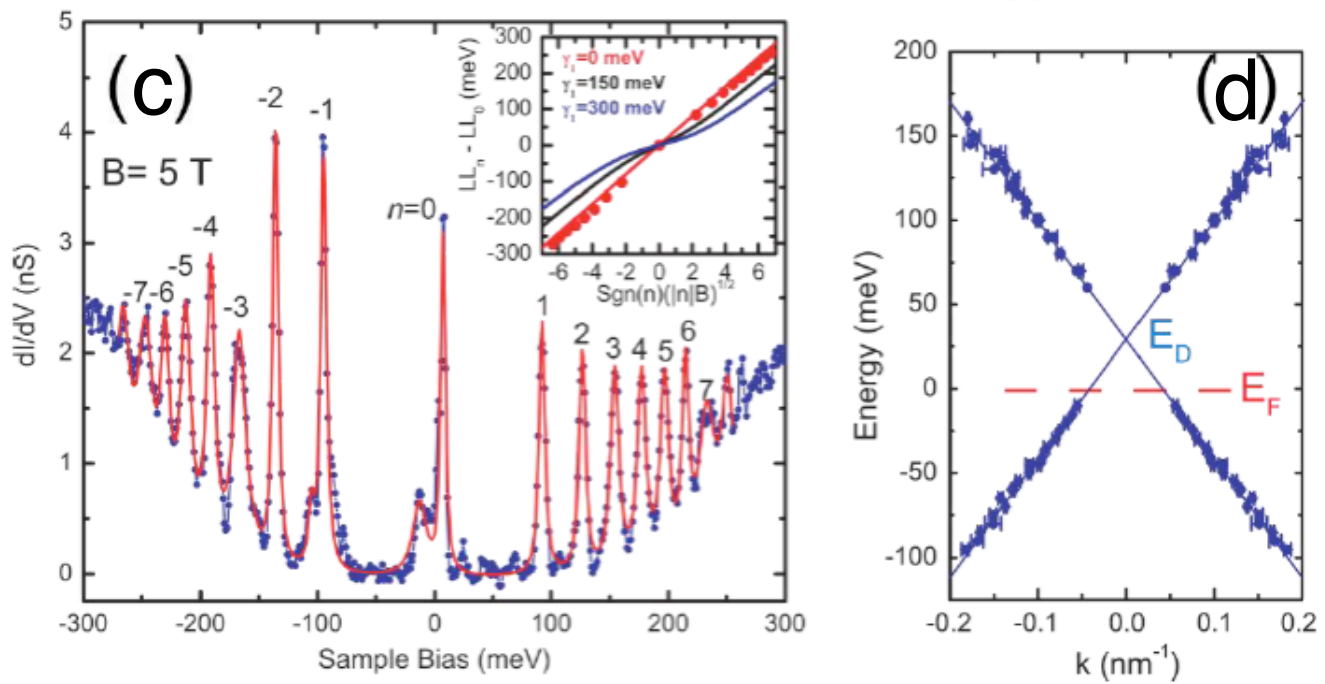

Figure 2

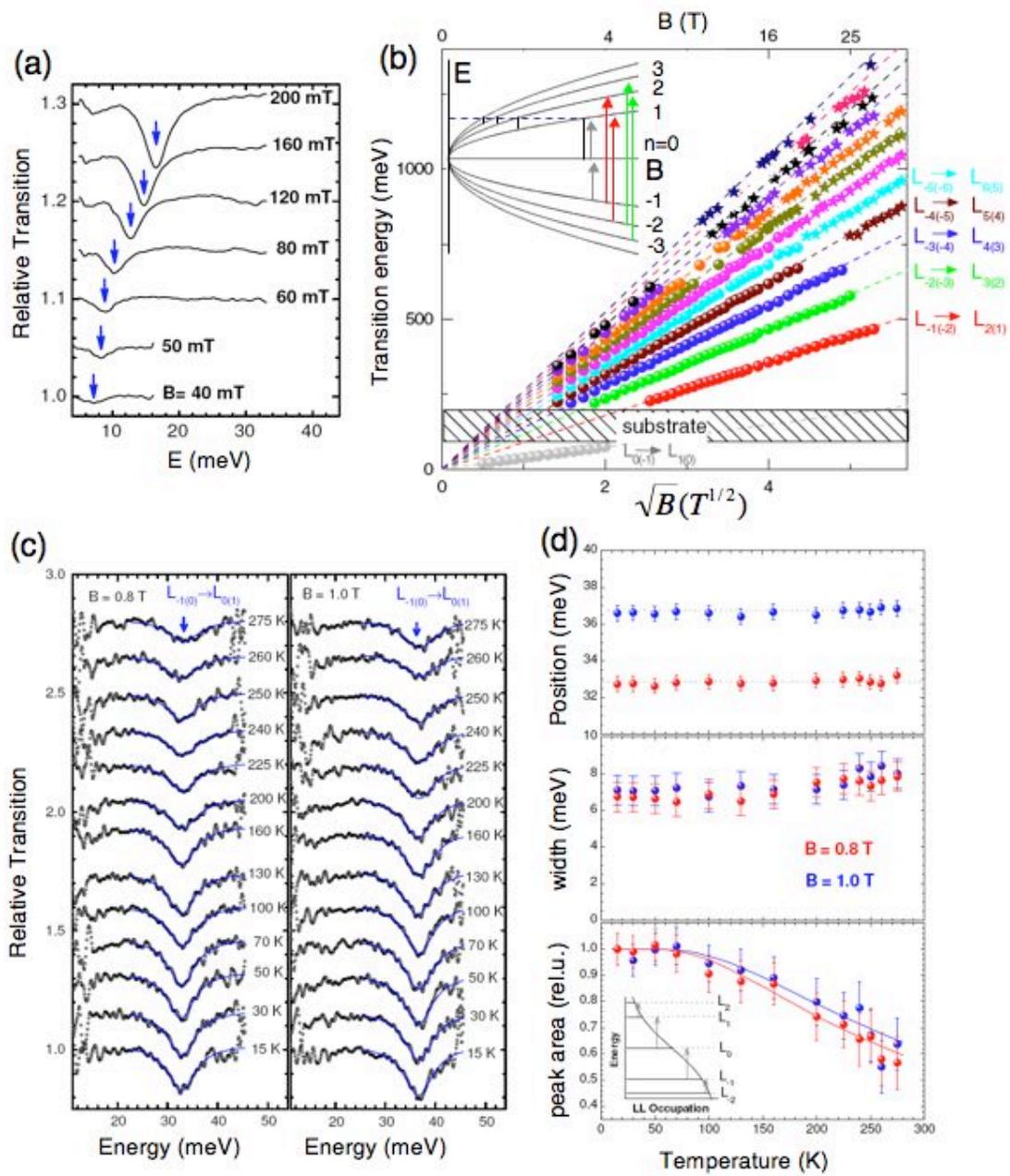

Figure 3

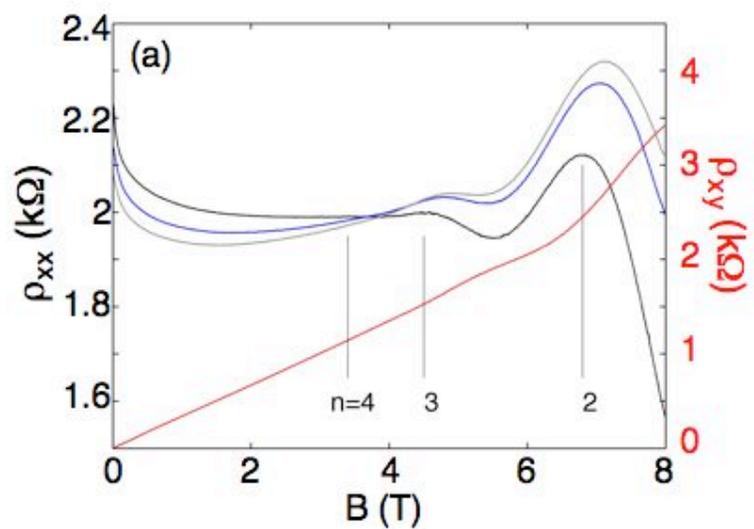
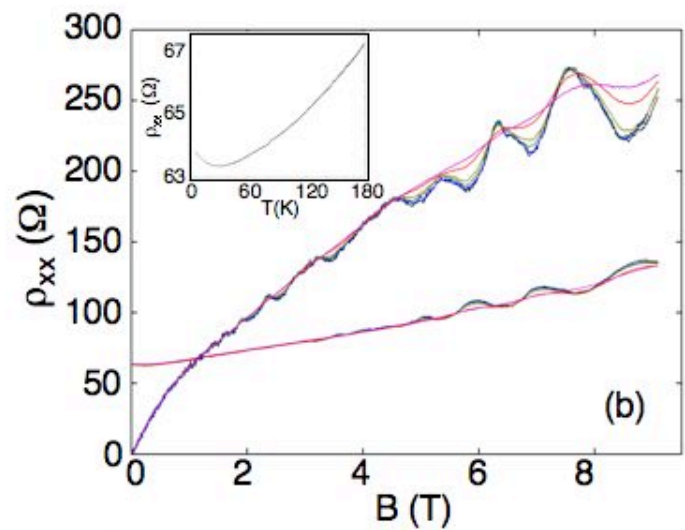
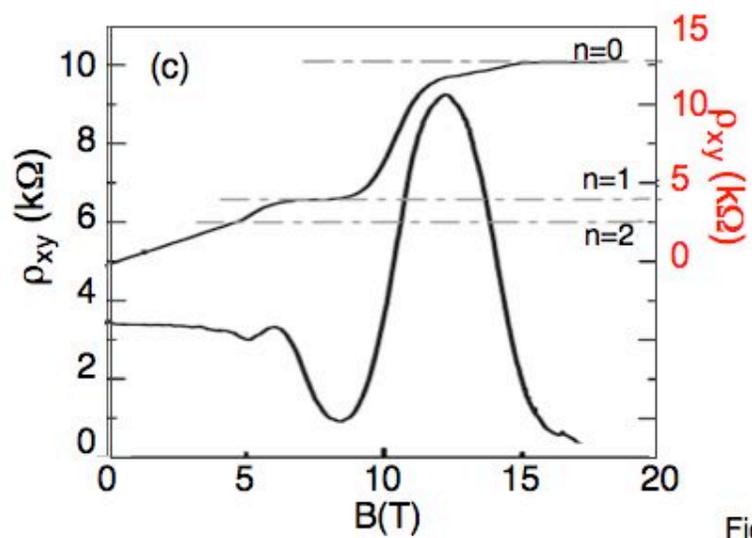
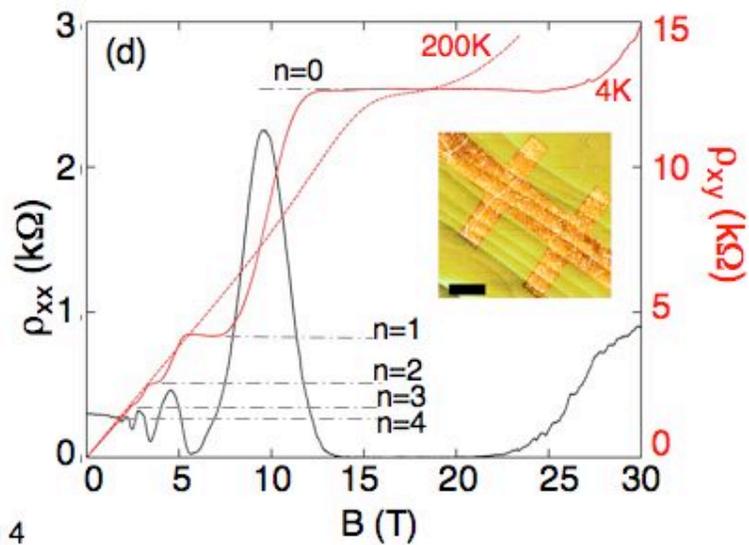

Figure 4

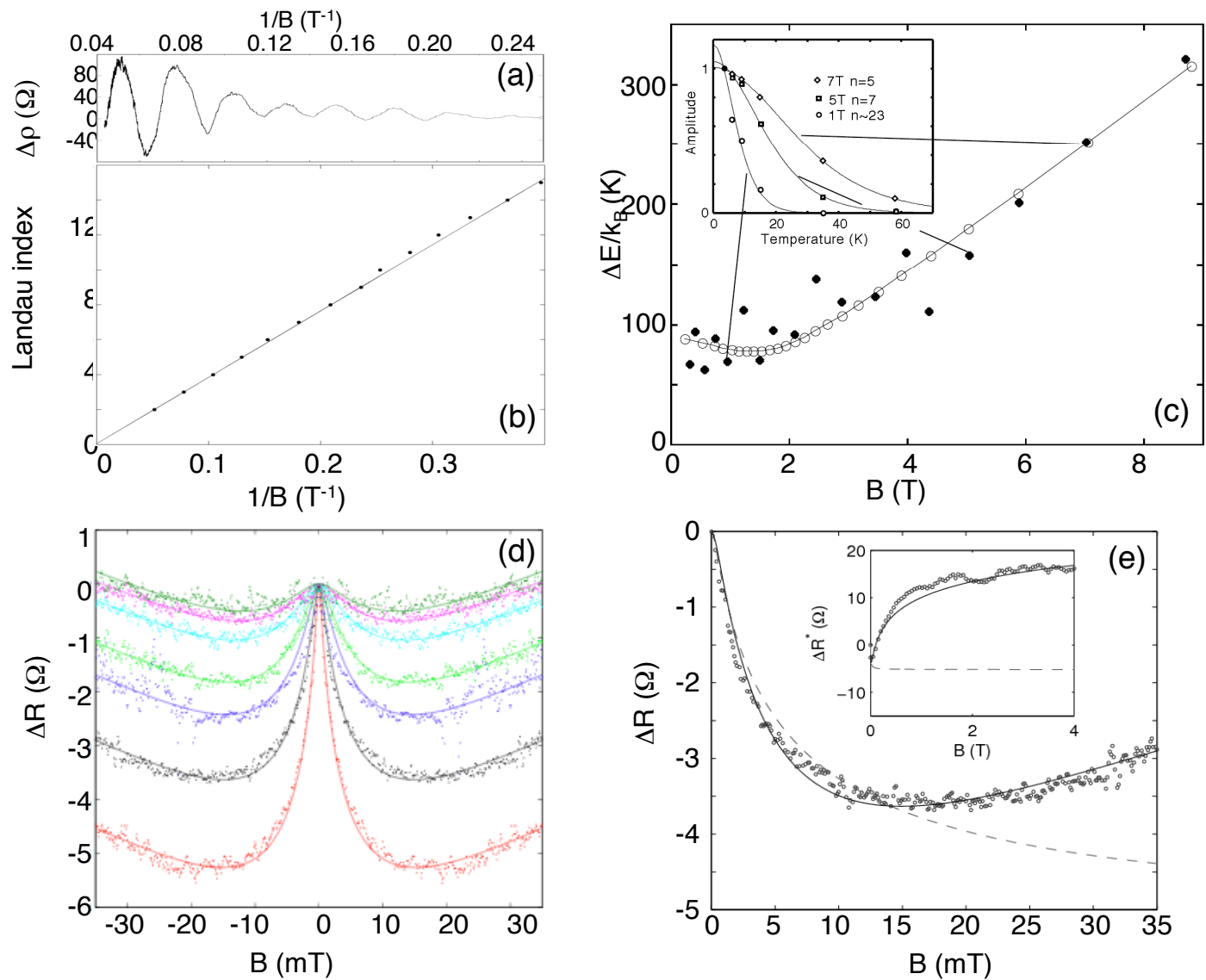

Figure 5

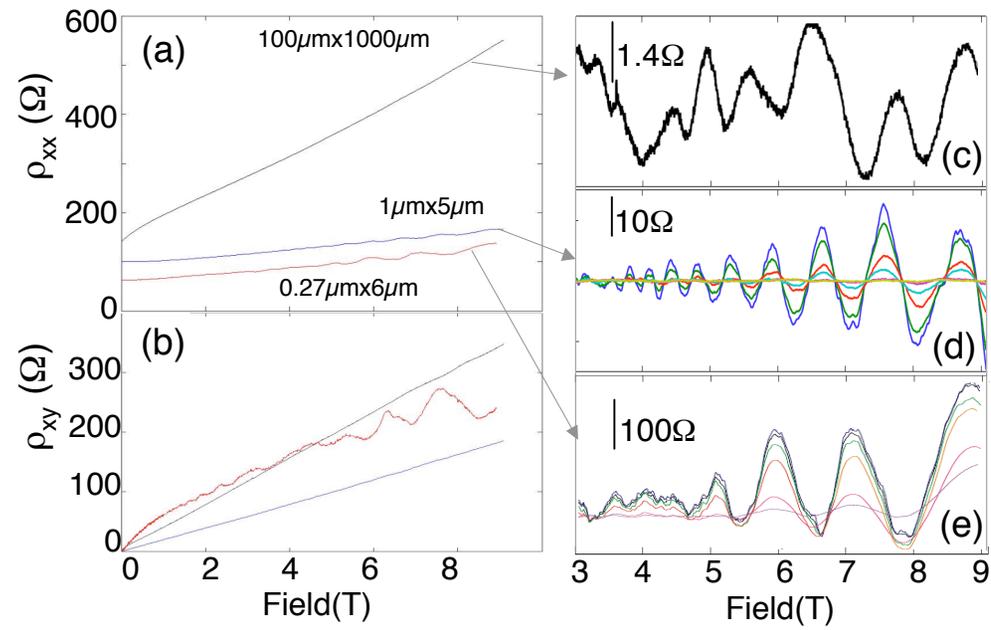

Figure 6

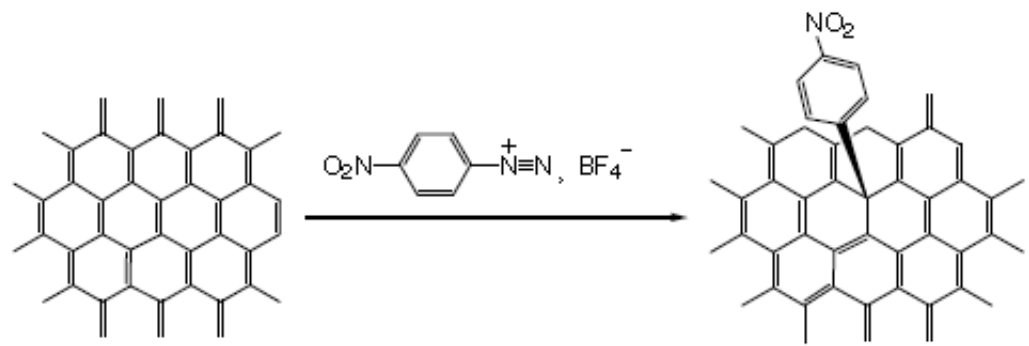

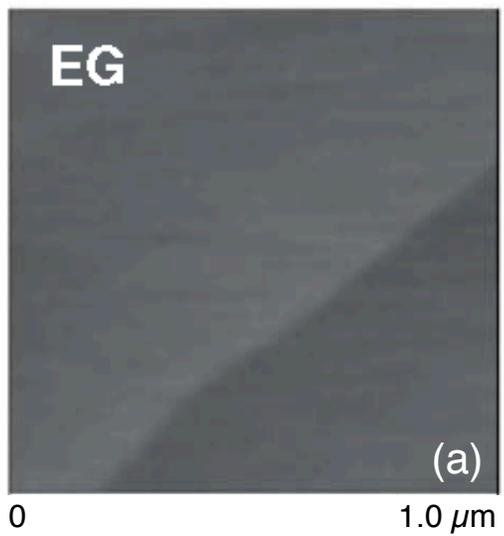
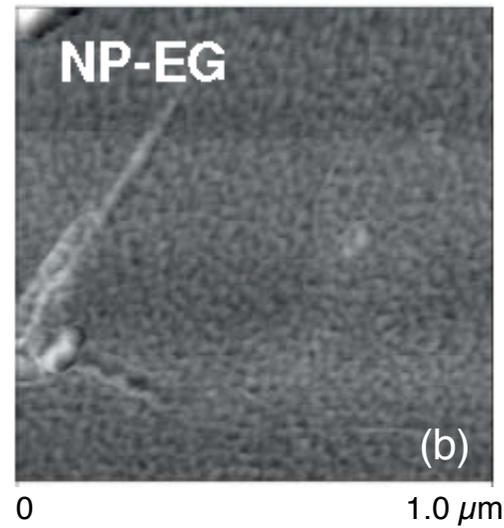
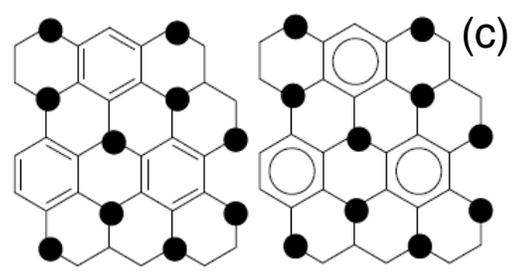
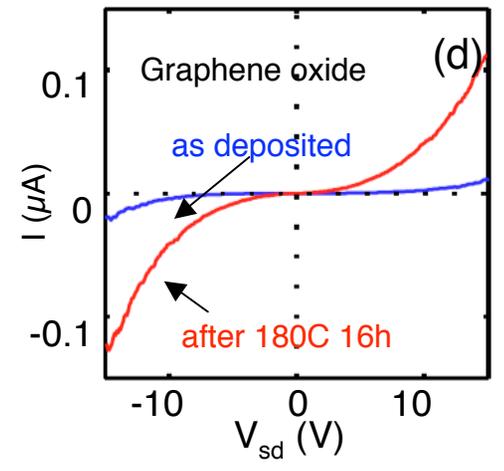

Figure 7

Figure 8

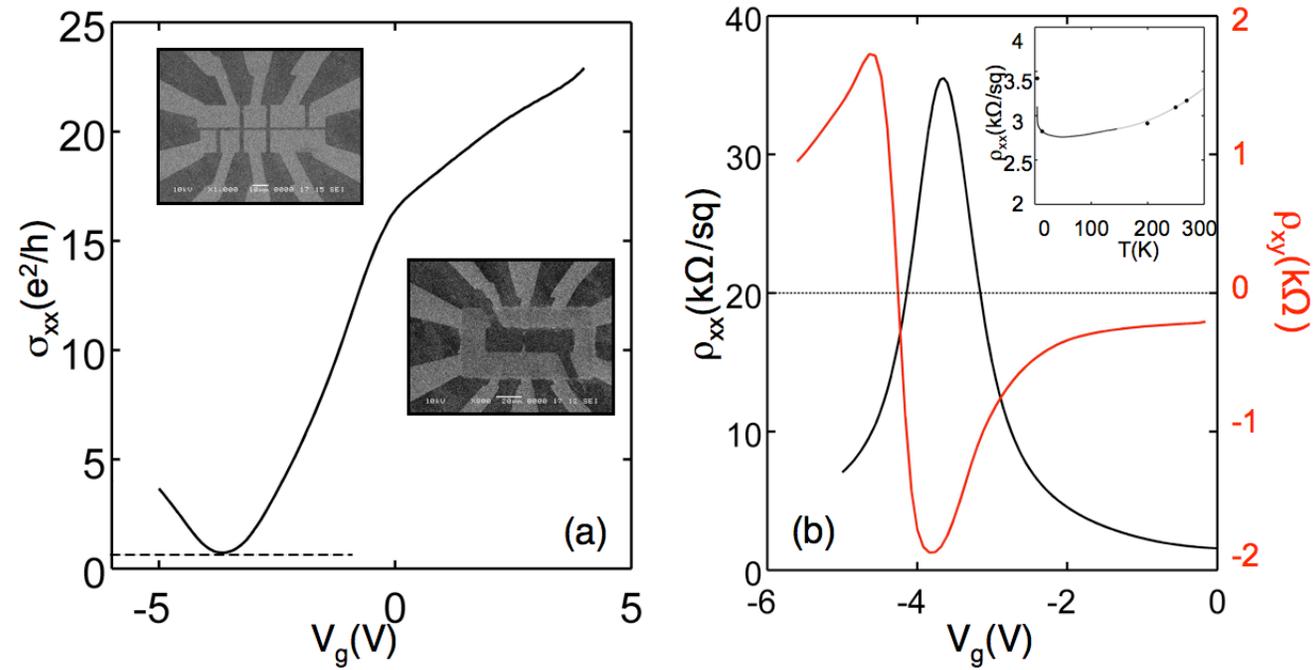
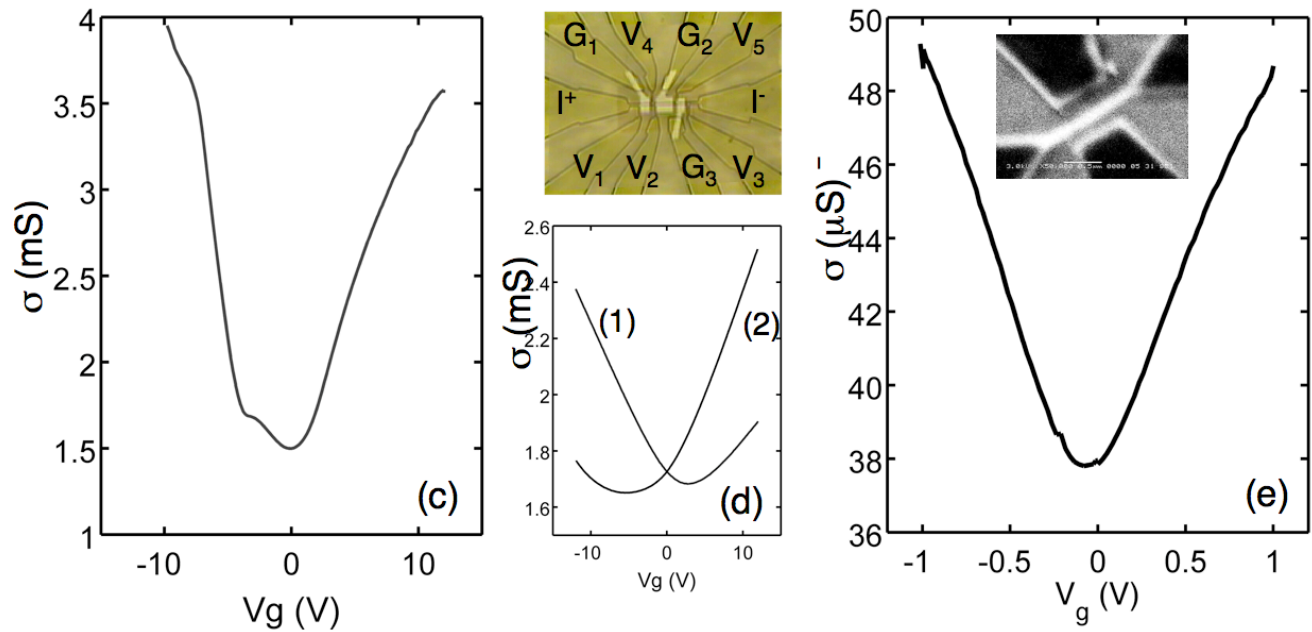

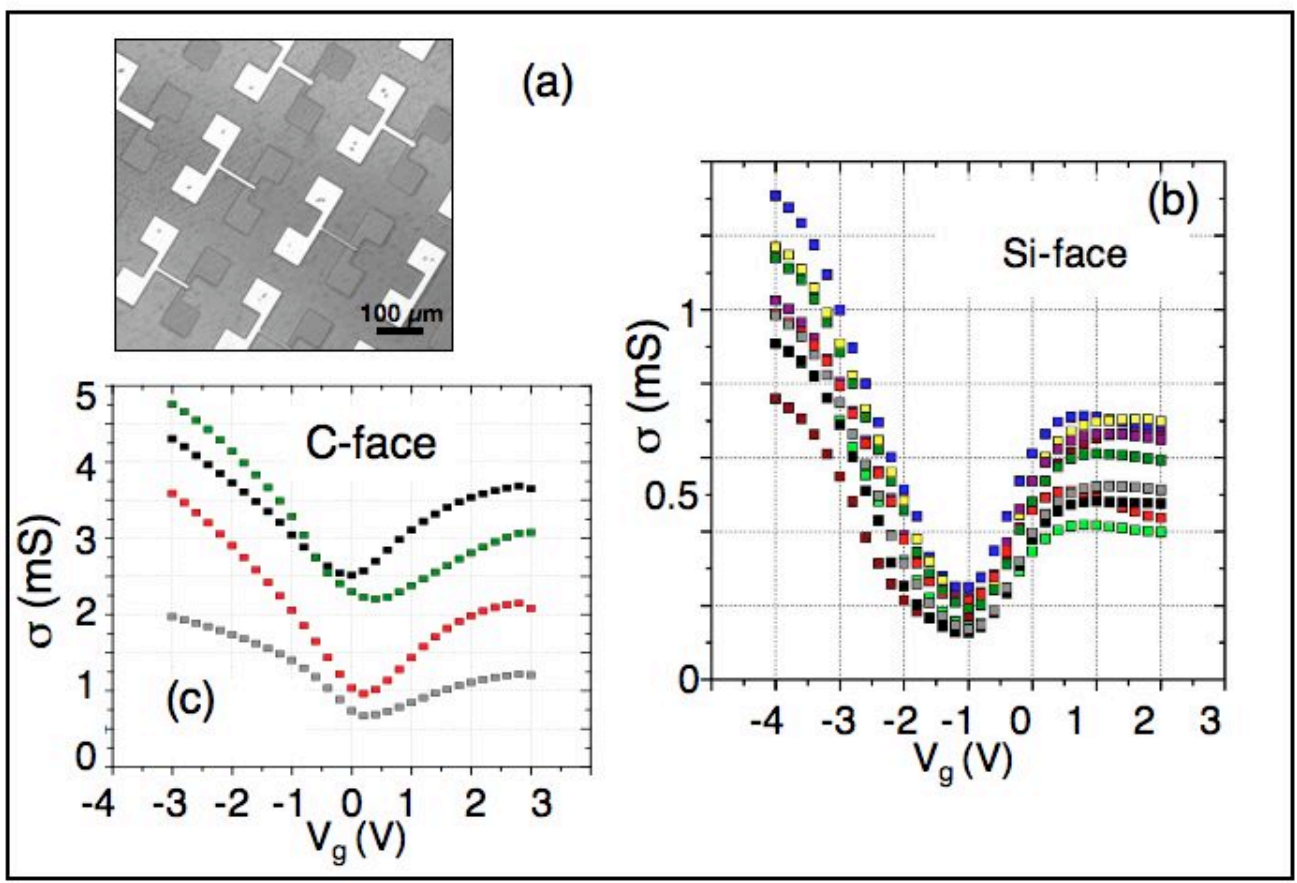

Figure 9

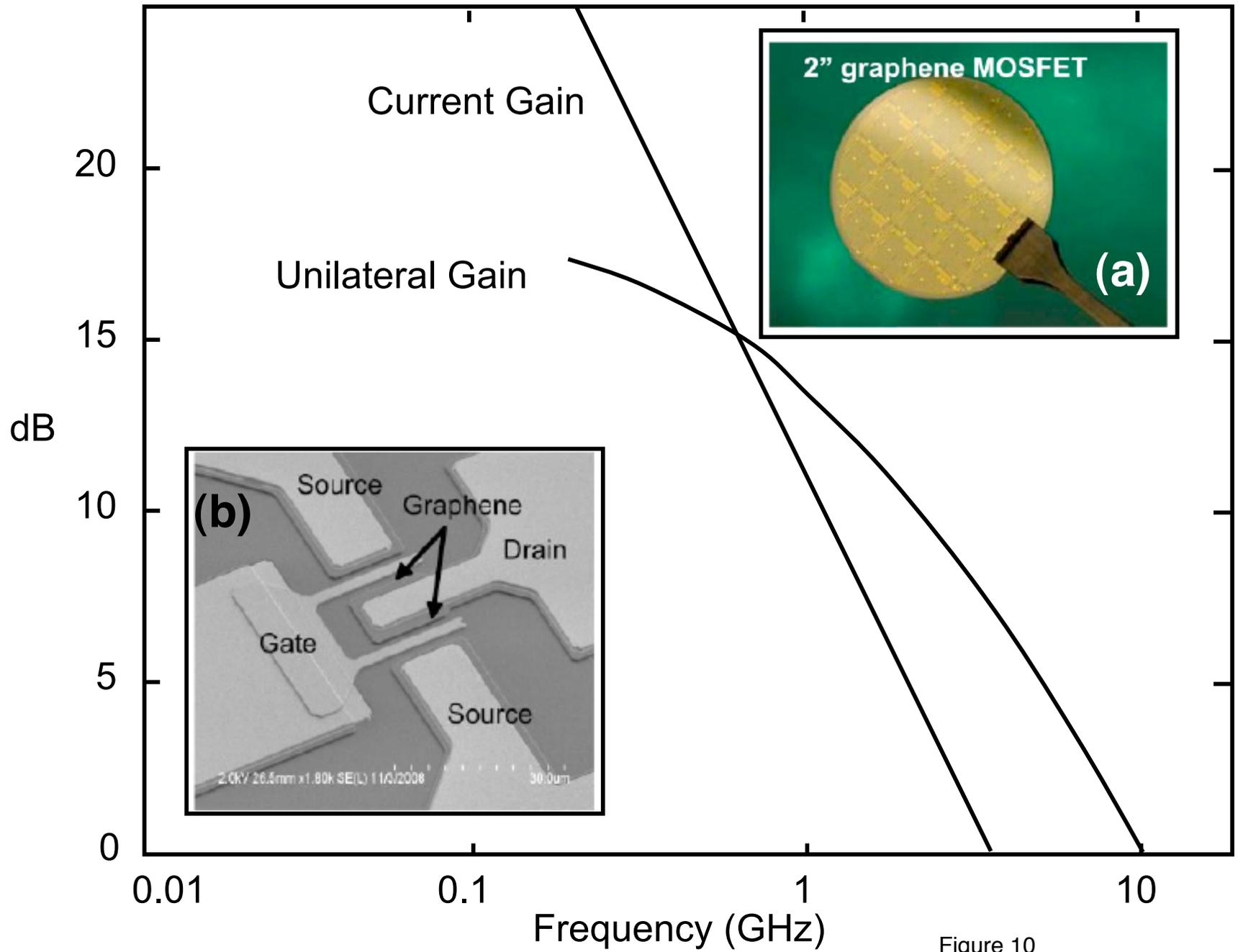

Figure 10